\documentclass[]{aa}
\usepackage{graphicx}
\begin{document}

\title{On the detection of chemically peculiar stars using $\Delta a$ photometry}
\author{E.~Paunzen\inst{1}, Ch.~St{\"u}tz\inst{1}, H.M.~Maitzen\inst{1}}

\mail{Ernst.Paunzen@univie.ac.at}

\institute{Institut f{\"u}r Astronomie der Universit{\"a}t Wien,
           T{\"u}rkenschanzstr. 17, A-1180 Wien, Austria}

\date{Received 7 March 2005 / Accepted 3 June 2005}
\titlerunning{}{}

\abstract{We have summarized all $\Delta a$ measurements for galactic field stars (1474 objects)
from the literature published over more than two
decades. These measurements were, for the
first time, compiled and homogeneously analyzed. The $\Delta a$ intermediate 
band photometric system 
samples the depth of the 5200\AA\, flux depression by comparing the flux at the center
with the adjacent regions with bandwidths of 110\AA\, to 230\AA. Because it was slightly modified 
over the last three decades, we checked for 
systematic trends for the different measurements but found no correlations whatsoever. 
The $\Delta a$ photometric system is most suitable to detecting magnetic chemically
peculiar (CP) stars with high
efficiency, but is also capable of detecting a small percentage of 
non-magnetic CP objects. Furthermore, the groups of (metal-weak) $\lambda$ Bootis, 
as well as classical Be/shell stars, can be successfully investigated. 
In addition, we also analyzed the behaviour of supergiants (luminosity class I and II).
On the basis of apparent normal type objects, the correlation of the 3$\sigma$ 
significance limit and
the percentage of positive detection for all groups was derived. 
We compared the capability of the $\Delta a$ photometric system with the
$\Delta (V1 - G)$ and $Z$ indices of the Geneva 7-color system to detect peculiar objects.
Both photometric systems show the same efficiency for the detection of CP 
and $\lambda$ Bootis stars, while the
indices in the Geneva system are even more efficient at detecting
Be/shell objects.
On the basis of this statistical analysis it is possible to derive the incidence of CP
stars in galactic open cluster and extragalactic systems including the former unknown bias
of undetected objects. This is especially important in order to make a sound statistical
analysis of the correlation between the occurrence of these objects and astrophysical parameters
such as the age, metallicity, and strength of global, as well as local, magnetic fields.
}
\maketitle

\section{Introduction}

The classical chemically peculiar (CP) stars of the upper main sequence (luminosity
classes V and IV) are
targets of detailed investigations since their ``discovery'' by
Maury (1897). They are excellent test objects for astrophysical 
processes like diffusion, convection, and stratification in stellar
atmospheres in the presence of rather strong magnetic fields. These mechanism 
can be studied depending on the age and metallicity of individual objects,
which was a major setback of the stellar evolution theory.

There is a wide variety of peculiar stars in the spectral domain from
B0 (30000\,K) to F5 (6500\,K). Preston (1974) divided the CP stars into four
different groups selected on the basis of the presence of strong magnetic fields
and the kind of the surface elemental peculiarity. Later on, small peculiar groups like
the $\lambda$ Bootis stars were also investigated (Paunzen et al. 2002a) and
classified as an individual subgroup.

The prerequisite for investigating larger samples of CP stars (including the
generally fainter open cluster members) is an unambiguous detection. Looking into
catalogues of CP stars, especially the magnetic ones, it immediately becomes
obvious that there are many discrepancies even at classification dispersions.
The reasons for discrepant peculiarity assessments are mainly to be found in the
differences in observing material (density of
spectrograms, widening of spectra, dispersion, focussing), 
but also intrinsic variability of peculiar spectral features (e.g. silicon lines).

Besides the use of (very time consuming and magnitude limited) high dispersion spectroscopy,
photometry has shown a way out of this dilemma, especially through
the discovery of characteristic broad band absorption features,
the most suitable of them located around 5200\AA.
Kodaira (1969) was the first notice to significant flux depressions at
4100\AA, 5200\AA, and 6300\AA\, in the spectrum of the CP star HD~221568.
These flux depressions are very likely enhanced by the
effects of magnetic radiative transport phenomena. 

Nearly three decades ago, Maitzen (1976) introduced the narrow-band, three filter 
$\Delta a$ photometric system in order to investigate the flux depression at 5200\AA, 
which samples the depth of this feature
by comparing the flux at the center (5220\AA, $g_{\rm 2}$)
with the adjacent regions (5010\AA, $g_{\rm 1}$ and 5500\AA,
$y$) using bandwidths of 110\AA\, to 230\AA. The respective index was introduced as:
$$ a = g_{\rm 2} - (g_{\rm 1} + y)/2. $$
Since this quantity is slightly dependent on the temperature (increasing
towards lower temperatures), the intrinsic peculiarity index
had to be defined as
$$ \Delta a = a - a_{\rm 0}[(b - y); (B - V); (g_{\rm 1} - y)], $$
i.e. the difference between the individual $a$-values and those
of non-peculiar stars for the same color. The locus of the $a_{\rm 0}$-values
has been called the normality line.

\begin{table}[t]
\begin{center}
\caption{The number of individual $\Delta a$ measurements from the given reference. The used
filter systems are according to Table \ref{filters}.}
\label{n_data}
\begin{tabular}{lrc}
\hline
\hline
Ref. & $N_{obs.}$ & System \\
\hline
Maitzen (1976) & 168 & 1 \\
Maitzen (1980a) & 10 & 2 \\
Maitzen (1980b) & 8 & 2 \\
Maitzen \& Seggewiss (1980) & 21 & 2 \\ 
Maitzen (1981) & 20 & 2 \\ 
Maitzen \& Vogt (1983) & 342 & 3 \\
Maitzen \& Pavlovski (1989a) & 16 & 4 \\ 
Maitzen \& Pavlovski (1989b) & 31 & 4 \\
Pavlovski \& Maitzen (1989) & 40 & 4 \\
Schnell \& Maitzen (1994) & 3 & 5 \\
Schnell \& Maitzen (1995) & 14 & 5 \\
Maitzen et al. (1997) & 6 & 6 \\
Maitzen et al. (1998) & 131 & 7 \\
Vogt et al. (1998) & 803 & 3 \\
\hline
\end{tabular}
\end{center}
\end{table}

Since its introduction in 1976, many papers presenting $\Delta a$ photometry of
galactic field and especially open clusters have been published. Until the late 90s, all 
observations were performed with photomultipliers. 

All but one (Vogt et al. 1998) publication about $\Delta a$ photometry of galactic
field stars were explicitly devoted to chemically peculiar stars or related
groups selected on the basis of several relevant catalogues. Superficial ``normal'' type
objects were only observed in order to define the normality line for the specific set of
data.

For the first time, we have combined all available $\Delta a$ data of bright galactic
field stars that have been published in the literature over more than
two decades, in order to make a sound statistical analysis
of detection probabilities for all kinds of peculiar objects, as well as 
luminosity class I/II supergiants. This is especially important for observations
in open clusters (Paunzen et al. 2002b) and even in the Magellanic clouds (Maitzen et al.
2001). The knowledge of how many peculiar stars for a certain detection limit are
expected in comparison with the actually observed number, serves as important information
for a statistical analysis, such as the incidence of CP stars that depend on different
local environments. 

\section{Selection, preparation, and homgenization of the data}

We used the following sources of $\Delta a$ photometry for
galactic field stars: Maitzen (1976, 1980a, 1980b, 
1981), Maitzen \& Seggewiss (1980), Maitzen \& Vogt (1983), 
Maitzen \& Pavlovski (1989a, 1989b), Pavlovski \& Maitzen (1989),
Schnell \& Maitzen (1994, 1995), Maitzen et al. (1997,
1998) and Vogt et al. (1998). The number of individual measurements are given in
Table \ref{n_data}. Some references already list $(b-y)_0$ magnitudes, these were
used for this study. Objects were only included which have Str{\"o}mgren indices
available.

All but Maitzen et al. (1997; CCD) were conducted with
photomultipliers and classical aperture photometry (one star at one time). The different
filter sets of these observations are listed in Table \ref{filters}. The only important difference
is the filter set used by Maitzen et al. (1997) together with the CCD equipment.
This new CCD photometric $\Delta a$ system was successfully applied to open clusters
of the Milky Way (Paunzen et al. 2002b) and the Large Magellanic Cloud (Maitzen et al. 2001).
Although the bandwidths of $g_1$ and $y$ were exchanged, the overall values of
$\Delta a$ are in very good agreement with those of the photoelectric system.

\begin{table}[t]
\begin{center}
\caption{Filter systems used for the $\Delta a$ measurements.}
\label{filters}
\begin{tabular}{ccccccc}
\hline
\hline
System & \multicolumn{2}{c}{$g_1$} & \multicolumn{2}{c}{$g_2$} & \multicolumn{2}{c}{$y$} \\
     & $\lambda_c$ & $\lambda_{1/2}$ & $\lambda_c$ & $\lambda_{1/2}$ & $\lambda_c$ 
	 & $\lambda_{1/2}$ \\
\hline
1    & 5020 & 130 & 5240 & 130 & 5485 & 230 \\
2    & 5010 & 130 & 5215 & 130 & 5485 & 230 \\
3    & 5020 & 130 & 5240 & 130 & 5505 & 230 \\
4    & 5026 & 111 & 5238 & 138 & 5466 & 230 \\
5    & 5017 & 120 & 5212 & 120 & 5494 & 228 \\
6    & 5027 & 222 & 5205 & 107 & 5509 & 120 \\
7    & 5017 & 110 & 5212 & 120 & 5473 & 188 \\
\hline
\end{tabular}
\end{center}
\end{table}

The Johnson, Geneva, and Str{\"o}mgren (if not listed
in the original source of the $\Delta a$ photometry) colors 
were taken from the General Catalogue of 
Photometric Data (GCPD, Mermilliod et al. 1997). 
Usually, the reddening for objects within the solar neighborhood is estimated
using photometric calibrations in the Str{\"o}mgren $uvby\beta$ system
(Str{\"o}mgren 1966; Crawford \& Mander 1966; Crawford 1975, 1979; Hilditch et al. 1983).
The validity of these procedures for chemically peculiar stars was
shown, for example, in Maitzen et al. (2000).
However, these relations have to be used with some caution when applied 
to magnetic CP2 stars, because all 
calibrations are primarily based on the $\beta$ index. But due to variable 
strong magnetic fields, the $\beta$ index can give erratic values 
(Catalano\& Leone 1994). Another effect taken into account is
a ``blueing'' effect, which occurs in bluer colors 
due to stronger UV absorption than in normal type stars, which can be 
erratically interpreted as strong reddening (Adelman 1980).
An independent way to derive the interstellar reddening is to use galactic
reddening maps, which are derived from open clusters as well as from galactic
field stars. For this comparison, 
we used the model proposed by Chen et
al. (1998) to derive the interstellar reddening for all program stars. The
values from the calibration of the Str{\"o}mgren $uvby\beta$ and the model by
Chen et al. (1998) agree within an error of the mean of $\pm$1.9\,mmag
for the complete sample. The normality line is 
shifted by $E(g_{\rm 1} - y)$\,=\,0.55$E(b-y)$ 
to the red and by a small amount $E(a)$ to higher $a$-values (Maitzen 1993)
$$a({\rm corr}) = a({\rm obs}) - 0.07E(b - y).$$

In order to check this procedure, we used the raw data of Maitzen (1976)
to derive the normality line of these data. He used a combined
normality line for the $(b-y)$ and $(b-y)_0$ values under the assumption
that the reddening in the solar neighborhood is negligible. Our results 
confirm his $A_0$ parameter with a slight adjustment of the $A_2$ value from
0.06 to 0.066, which means a correction of $\Delta a$ in the order of only one 
mmag. This leads to confidence in our dereddening procedure.    

As the next step, we checked the intrinsic consistency of the different
reference for objects in common since the filter systems 
are slightly different (Table \ref{filters}).
Besides a possibly wrong identification or typographical errors, 
the variability of $\Delta a$
itself has to be taken into account. Several chemically peculiar stars
show variability in correlation with strong magnetic fields and rotation. These
variations were also detected in the $\Delta a$ photometric system (Maitzen \& Vogt
1983, Fig. 11). There is also the outstanding case of Pleione (Maitzen \& Pavlovski
1987). This Be/shell star shows a strong positive $\Delta a$ value (+36 mmag) as it
undergoes its shell phase. 

The published $(b-y)_0$ and $\Delta a$ values were checked in this respect and no significant
trends were found. All references agree, in a statistical sense, within one
sigma of the correlation coefficients. We found no significant outliers.
Therefore, the values of objects with more than one measurement were averaged
without using any weights,
resulting in 1474 individual galactic field stars (which almost doubles the sample
used by Vogt et al. 1998) with an available $\Delta a$
value. This sample consists of ``normal'' type and peculiar stars of all kinds.

\begin{figure}[t]
\begin{center}
\includegraphics[width=75mm]{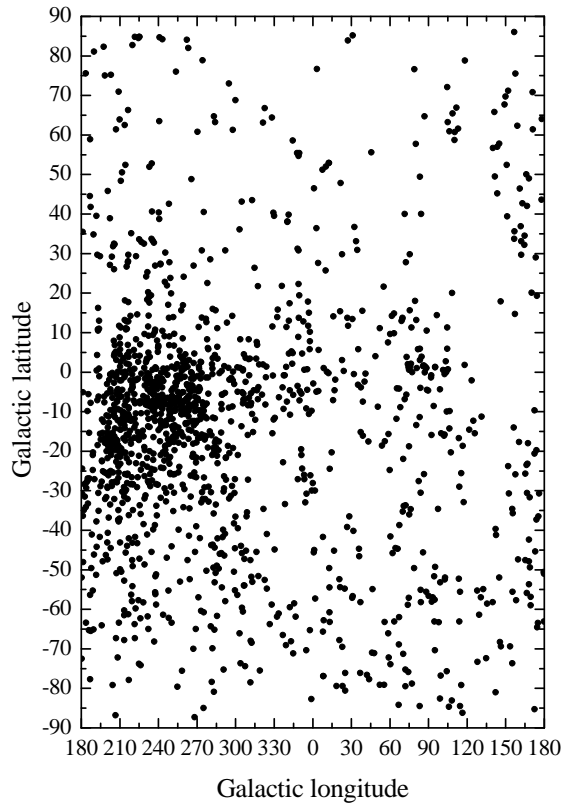}
\caption{The galactic distribution of our sample including 1474 individual objects.
There is a clustering of objects at 190$^o$\,$<$\,$l$\,$<$\,280$^o$, the
almost reddening free (up to 500\,pc from the Sun) third galactic quadrant. }
\label{galactic}
\end{center}
\end{figure}

\begin{figure*}
\begin{center}
\includegraphics[width=150mm]{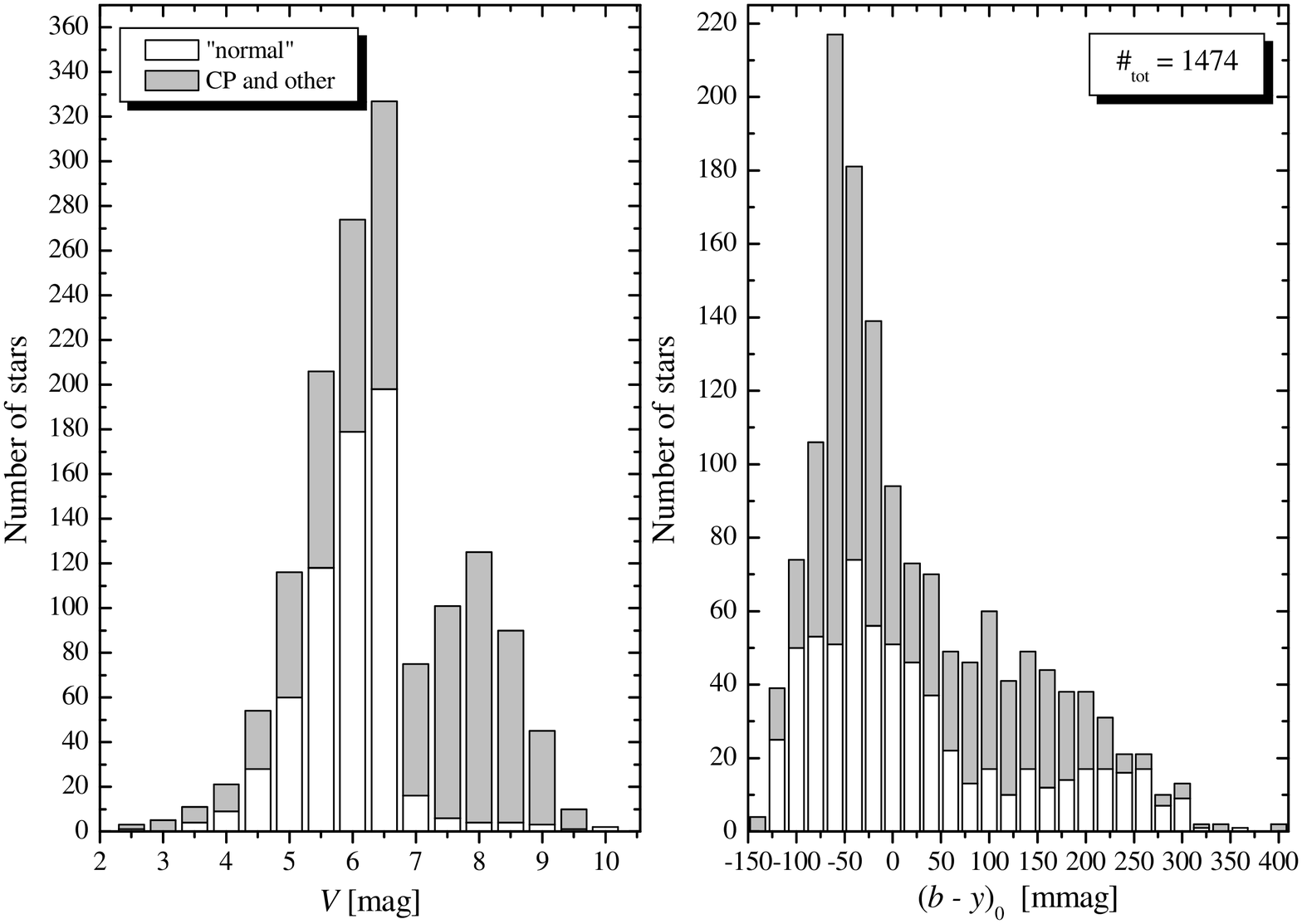}
\caption{The distribution of Johnson $V$ and Str{\"o}mgren $(b-y)_0$ of
the  sample divided into ``normal'' type and ``other'' (all objects included in
Renson 1991, Table \ref{limits} and Sect. \ref{out}) stars.}
\label{histos}
\end{center}
\end{figure*}

\begin{figure}
\begin{center}
\includegraphics[width=65mm]{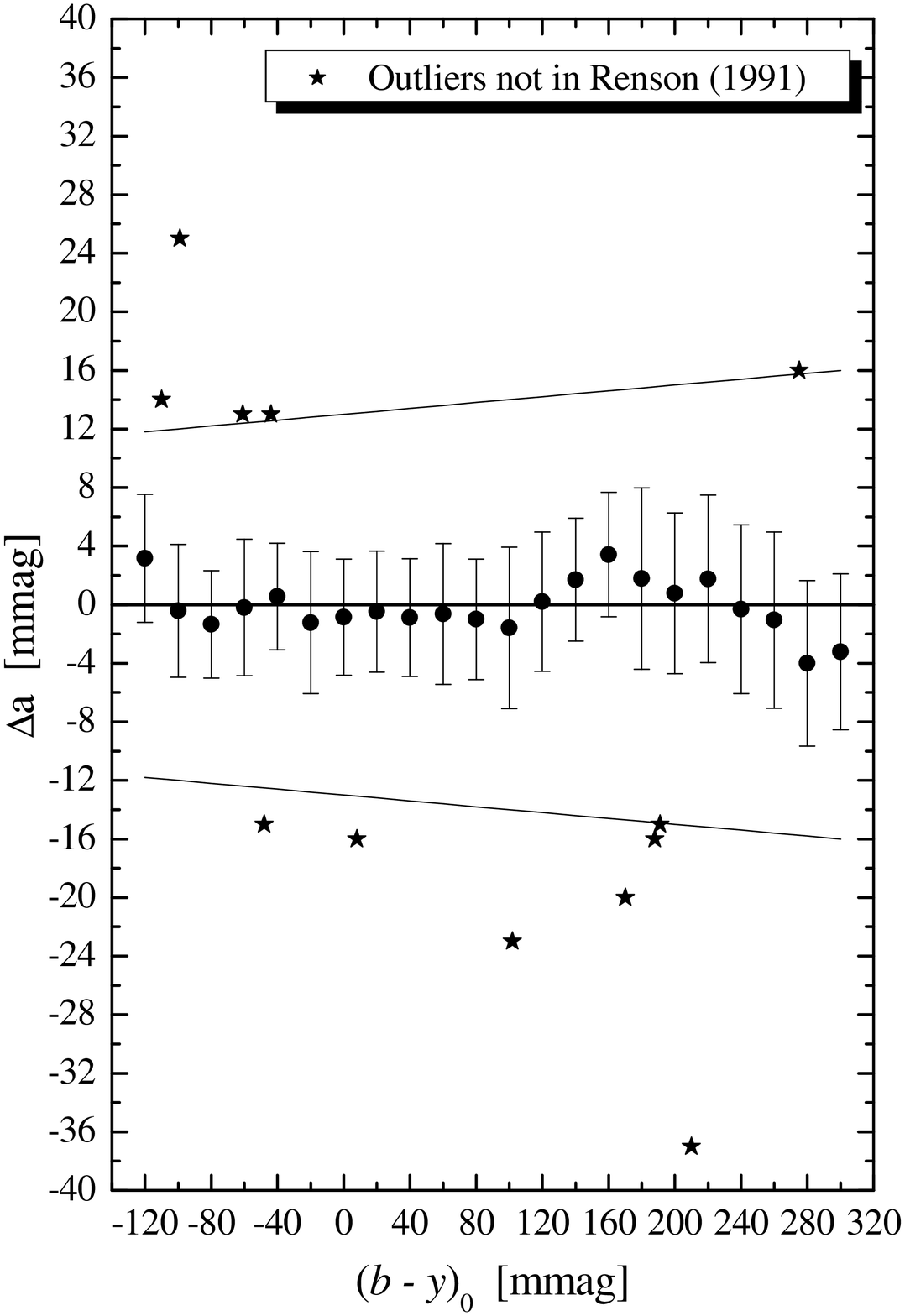}
\caption{The mean values and
the corresponding root mean square scatter for all normal type objects
with $\Delta a$ measurements. The straight lines are the 3$\sigma$ fits
through the errors, whereas the asterisks denote the outliers of objects
not included in the catalogue of Renson (1991) discussed in Sect. \ref{out}.}
\label{flimit}
\end{center}
\end{figure}

\section{Analyzing the sample} \label{sample}

With the sample of 1474 galactic field stars, a statistically sound
analysis was performed.
Figure \ref{galactic} shows the distribution of the galactic coordinates
for the sample. There is a clustering of objects between 
190$^o$\,$<$\,$l$\,$<$\,280$^o$ (third galactic quadrant) because of the systematic investigation 
in the southern hemisphere by Vogt et al. (1998). In this region, the
reddening is almost negligible for distances up to 500\,pc (Chen et al. 1998).
But no unknown bias has been introduced, because the chemically peculiar 
stars are distributed uniformly in the solar neighborhood (G\'omez et al. 1998). 

The sample covers $-$0.15\,$<$\,$(b-y)_0$\,$<$\,0.4\,mag (Fig. \ref{histos}),
which corresponds to the spectral range from B0 to F8 (Crawford 1975, 1978).
The peak of the distribution is at $(b-y)_0$\,=\,$-$0.05\,mag or B8 at the 
main sequence. At this spectral type, the number of CP2 stars reaches an
intrinsic
maximum. The Johnson $V$ distribution shows two maxima around $V$\,=\,6\,mag
and 8\,mag and extends from 1.5\,$<$\,$V$\,$<$\,10.5\,mag. The first peak is
a consequence of the systematical investigation of all bright stars by
Vogt et al. (1998) whereas the second one is due to the chosen sample
by Maitzen \& Vogt (1983). They have investigated the list of astrophysically
interesting stars by Bidelman \& MacConnell (1973), which was published prior
to the Michigan spectral survey, while both are based on the same 
spectroscopic material.

There are three stars in the sample that exhibit positive $\Delta a$
values, for which we were not able to designate a distinctive
membership
in a certain class of CP stars. Certainly, these three objects deserve
further interest: \\
{\it HD 68161:} Eggen (1980) lists a photometric spectral type of B7III, whereas the 
spectroscopic type is B8Ib/II (Houk \& Swift 1999). This is a strong
indication that this star is of CP2 type, because the silicon lines are also
used to derive the luminosity classification. \\
{\it HD 202671:} classified as B6(8)III SrTi He-w (Bychkov et al. 2003), it
seems to be a transition object of the CP3 and CP4 group. \\ 
{\it HD 225253:} we found several deviating classifications in the literature,
B8III (Jaschek et al. 1969), B8IV/V p(mild) (Maitzen 1980a), and B7 UV gallium 
(Jaschek \& Jaschek 1987). With a $\Delta a$ value of +18 mmag, this object
is obviously chemically peculiar. \\

\begin{table}[t]
\begin{center}
\caption{Limits in mmag for detecting the individual groups of objects. For example
49\% of all CP1 stars can be detected with a 3$\sigma$ limit of +5 mmag. In
addition, the mean group $\overline{\Delta a}$, the maximum values for group
members $\Delta a_{m}$, and the number $N$ of well-established members for the
analysis are listed.}
\label{limits}
\begin{tabular}{lrrrr|rrr}
\hline
\hline
\vspace*{-3mm}
\\
& \multicolumn{1}{c}{+5} & \multicolumn{1}{c}{+10} & \multicolumn{1}{c}{+15} & 
\multicolumn{1}{c}{+20} & \multicolumn{1}{c}{$\overline{\Delta a}$} &  
\multicolumn{1}{c}{$\Delta a_{m}$} & \multicolumn{1}{c}{$N$} \\
\hline
CP1 & 49\% & 17\% & 8\% & 1\% & +3.2 & +22 & 78 \\
CP2 & 98\% & 93\% & 90\% & 79\% & +32.5 & +79 & 108 \\
CP3 & 80\% & 10\% & $-$ & $-$ & +5.3 & +11 & 10 \\
CP4 & 94\% & 94\% & 77\% & 35\% & +18.8 & +36 & 17 \\
I/II & 61\% & 28\% & 17\% & 11\% & +5.4 & +19 & 18 \\
\hline
& \multicolumn{1}{c}{$-$5} & \multicolumn{1}{c}{$-$10} & \multicolumn{1}{c}{$-$15} & 
\multicolumn{1}{c}{$-$20} \\
\hline
$\lambda$ Boo & 95\% & 65\% & 35\% & 20\% & $-$16.2 & $-$35 & 20 \\
Be & 41\% & 12\% & 5\% & $-$ & $-$1.5 & $-$19 & 59 \\
\hline
\end{tabular}
\end{center}
\end{table}

\subsection{The normal type objects}

The selection of apparent normal type objects is a rather difficult
task because of undetected peculiarities of all kinds which might introduce
an unknown bias. We chose all objects not listed
in the catalogue by Renson (1991), because he compiled stars which have been
identified as peculiar at least once in the literature, even though they
might turn out to be ``normal'' after all. As a next step, stars classified
as $\lambda$ Bootis (Paunzen et al. 2002a), Be/shell stars, and super giants of
luminosity classes I and II were excluded. For this purpose the spectral 
classifications given in the Michigan catalogues of two-dimensional 
spectral types (Houk \& Swift 1999 and references therein) and the extensive
list of Skiff (2003) were used. Known binary systems of all kinds were not
automatically excluded.

In total, 633 objects with luminosity classes V, IV, and III
were selected on this basis. These stars were divided
into subsamples according to their $(b-y)_0$ values with a bin size of 20\,mmag,
which corresponds to $\pm$1 spectral subclass.
For each subsample we calculated the mean $\Delta a$ value and
the corresponding root mean square scatter for all objects within it.
Figure \ref{flimit} shows the results,
with mean values uniformly distributed around zero. The straight lines
$\pm$13\,$\pm$\,0.01$(b-y)_0$
are linear fits through the mean values adding 3$\sigma$ in both directions.
The mean value of $\sigma$ is 4.8(8) mmag in the interval from  
$-$0.13\,$<$\,$(b-y)_0$\,$<$\,0.31\,mag. This is in line with the values
listed in Maitzen (1976) and Vogt et al. (1998). The detection limit for our
sample increases from $\pm$12 to $\pm$16 mmag towards cooler objects (Fig. \ref{flimit}).
There are two main reasons for this behaviour: 1) $g_1-y$ is no longer only an indicator
of the effective temperature because of significantly increasing 
line blanketing (like $b-y$) and 2) the strong general increase in metallic
lines towards later spectral types. Since most of the ``normal'' type 
stars are brighter
than $V$\,=\,7\,mag (Fig. \ref{histos}), we checked if the detection limit changes 
if only fainter ``normal'' objects are used. Although we are confronted with poor
number statistics, no correlation was found.

The observed scatter around the normality line has several causes.
Besides the instrumental limitations (e.g. photon noise and variable sky transparency),
which account for about 3\,mmag (Maitzen 1976),
the intrinsic variations due to the natural bandwidth of the main sequence
is important. Kupka et al. (2003) investigated a synthetic $\Delta a$ photometric
system based on modern stellar atmospheres and the filter transmission curves
of ``System 3'' in Table \ref{filters}. Their models are in the temperature range
from 7000 to 15000\,K and surface gravities from 2.5 to 4.5\,dex. They found a natural
bandwidth of about 4 mmag within these models. A metallicity
range from $-$0.5 to +0.5\,dex creates an overall bandwidth (= 2$\sigma$) of 10 mmag
(see their Figs. 3 and 4). This is more or less exactly the value we find for our
sample of normal type, galactic field stars. In open clusters for which the metallicity
is rather uniform, lower detection limits are expected if neglecting strong differential 
reddening. Paunzen et al. (2002b) reported 3$\sigma$ detection limits for five open
clusters between 7 and 9 mmag for objects with 9\,$<$\,$V$\,$<$\,18\,mag. Again,
this proves the high efficiency, as well as accuracy, of CCD $\Delta a$ photometry
in order to detect chemically peculiar objects.   

The results of the statistics for the different groups are listed in Table \ref{limits}.
The detection rates as functions of the corresponding 3$\sigma$ limits
are shown in Fig. \ref{glimits}.

\subsection{Outliers not included in Rensons catalogue} \label{out}

Twelve objects (five with positive and seven with negative $\Delta a$ values) 
are located outside the 3$\sigma$ limit (Fig. \ref{flimit}) and are not included in Renson (1991).
First of all, we discuss the seven objects with negative $\Delta a$ values
in more detail: \\
{\it HD 2834 ($(b-y)_0$\,=\,+8\,mmag, $\Delta a$\,=\,$-$16\,mmag):} a 
known spectroscopic binary system that is also part of 
a visible double system. \\
{\it HD 4158 (+210, $-$37):} a metal weak, cool star classified as hF3mF0V (wk met) 
by Paunzen (2001). \\
{\it HD 6173 (+102, $-$23):} although previously thought to be a member of the $\lambda$ Bootis
group, it was classified as A0IIIn by Paunzen et al. (2001). \\
{\it HD 38043 (+170, $-$20):} classified as blue giant by Norris et al. (1985). \\
{\it HD 97937 (+188, $-$16):} a metal weak (F0V m$-$1.5), intermediate Population
II type object (Gray 1989). \\
{\it HD 114911 ($-$48, $-$15):} a young, early type quadruple system which exhibits X-ray
emission (Hubrig et al. 2001). \\ 
{\it HD 217792 (+191, $-$15):} a Cepheid type variable within a spectroscopic binary
system. \\
The five objects with positive $\Delta a$ values lie only marginally (exception: HD 193237)
above the 3$\sigma$ limit (Fig. \ref{flimit}): \\
{\it HD 31726 ($-$110, +14):} a B1V star which exhibits rather normal silicon line
strengths (Massa 1989). \\
{\it HD 49028 ($-$61, +13):} a close visual binary system with a B7III and a F3V Fe$-$2
component (Corbally 1984). \\
{\it HD 65908 ($-$44, +13):} Bowyer et al. (1995) reported a clear detection in the
far-ultraviolet for this star which might indicate a previously unknown
binary nature. \\ 
{\it HD 73451 (+275, +16):} a known spectroscopic binary system with an early A type
and a G type component (Cowley et al. 1969). \\
{\it HD 193237, P Cygni, ($-$99, +25):} Rather 
outstanding, its true nature is still uncertain. Although classified as supernova,
there are strong indications that this object might have undergone a superoutburst typical
of luminous blue variables (Chu et al. 2004). This shows the potential of 
$\Delta a$ photometry to detect such objects which is especially important for studying
open clusters of the Milky Way and extragalactic systems.

\begin{table*}
\begin{center}
\caption{Confirmed chemically peculiar stars from Renson (1991), the type of peculiarity
was taken from the literature. The complete table is only available in electronic form.}
\label{cps}
{\tiny
\begin{tabular}{rrrrl|rrrrl|rrrrl}
\hline
\hline
\multicolumn{1}{c}{Renson} & \multicolumn{1}{c}{HD} & \multicolumn{1}{c}{$(b-y)_0$} & 
\multicolumn{1}{c}{$\Delta$a} & \multicolumn{1}{c}{Spec.} &
\multicolumn{1}{c}{Renson} & \multicolumn{1}{c}{HD} & \multicolumn{1}{c}{$(b-y)_0$} & 
\multicolumn{1}{c}{$\Delta$a} & \multicolumn{1}{c}{Spec.} &
\multicolumn{1}{c}{Renson} & \multicolumn{1}{c}{HD} & \multicolumn{1}{c}{$(b-y)_0$} & 
\multicolumn{1}{c}{$\Delta$a} & \multicolumn{1}{c}{Spec.} \\
\hline
30	&	315	&	$-$71	&	31	&	Si	&	15910	&	58292	&	$-$27	&	34	&	Si	&	28370	&	98486	&	$-$60	&	23	&	Si	\\
760	&	2957	&	$-$20	&	36	&	CrEu	&	16240	&	59435	&	284	&	25	&	SrCrSi	&	29270	&	101600	&	22	&	21	&	Si	\\
1480	&	5601	&	$-$56	&	49	&	Si	&	16550	&	60559	&	$-$62	&	23	&	Si	&	29330	&	101724	&	$-$78	&	16	&	Si	\\
1580	&	6164	&	$-$12	&	40	&	SiCrEu	&	16840	&	61382	&	$-$23	&	23	&	Si	&	29780	&	103302	&	4	&	33	&	SrCrEu	\\
1620	&	6322	&	$-$23	&	22	&	SrCrEu	&	17100	&	--	&	$-$19	&	46	&	Si	&	29820	&	103457	&	$-$7	&	35	&	Si	\\
1760	&	6783	&	$-$57	&	53	&	Si	&	17150	&	62530	&	$-$31	&	34	&	EuCr	&	29830	&	103498	&	$-$9	&	46	&	CrEuSr	\\
2110	&	8783	&	72	&	23	&	SrEuCr	&	17160	&	62535	&	$-$35	&	40	&	Si	&	30330	&	104810	&	$-$65	&	19	&	Si	\\
4060	&	16145	&	28	&	35	&	CrSrEu	&	17180	&	62556	&	35	&	18	&	EuCr	&	30460	&	105379	&	27	&	16	&	SrCr	\\
\hline
\end{tabular}
}
\end{center}
\end{table*}

\subsection{CP1 stars}

The Am/Fm stars (CP1) are preferably found within close binary systems. The main 
characteristics of this group are the lack of magnetic fields, the apparent underabundance
of calcium and scandium compared to the Sun, overabundances of Fe-peak elements,
and very low rotational velocities. Almost all CP1 stars seem to be rather evolved with ages
above 400 Myr (K{\"u}nzli \& North 1998). 

The observed abundance pattern is explained by the diffusion of elements together 
with the disappearance of the outer convection zone associated with
the helium ionization because of gravitional settling of helium (Michaud et al. 1983).
They predict a cut-off rotational velocity for such objects 
($\approx$\,90\,km\,s$^{-1}$), above which meridional circulation leads to a
mixing in the stellar atmosphere. 

In our sample, there are 78 well established CP1 stars yielding a slightly positive
mean value (+3.2 mmag) and quite extreme values (+22 mmag) for some members (e.g. HD
116235, HD 184552, and HD 204541). But the detection capability is only 17\% for a limit
of +10 mmag (Table \ref{limits}). Some of the most outstanding objects have already been
discussed by Vogt et al. (1998). Kupka et al. (2003) present detailed synthetic
$\Delta a$ values for the CP1 group concluding that the 5200\,\AA\, flux depression 
is only marginally detectable for these objects. This means that the most significant
elements contributing to this feature (e.g. Chromium) are not strongly enhanced
in CP1 stars.

Table \ref{cps} lists only one CP1 object with a significant positive $\Delta a$ value:
HD 196655, which was classified as probable Am star by Bidelman (1985). 

\subsection{CP2 stars}

This is the largest group of chemically peculiar stars already described by Maury (1897).
The main characteristics of the classical CP2 stars are: peculiar
and often variable line strengths, quadrature of line variability
with radial velocity changes, photometric variability with the same
periodicity, and coincidence of extrema. Slow rotation was inferred
from the sharpness of spectral lines. Overabundances of several orders
of magnitude compared to the Sun were derived for heavy elements
such as Silicon, Chromium, Strontium, and Europium. 

The strong global magnetic fields exhibit variability of the field strength 
including even a reversal of magnetic polarity leading
the Oblique Rotator concept of slowly rotating stars with non-coincidence
of the magnetic and rotational axes. This model produces
variability and reversals of the magnetic field strength similar to a 
lighthouse. Due to the chemical abundance concentrations at the magnetic
poles spectral and the related photometric variabilities are also
easily understood, as are radial velocity variations of the appearing
and receding patches on the stellar surface (Deutsch 1970).

Most of the $\Delta a$ observations were dedicated to this group because of
the high efficiency at detecting CP2 stars. This is also reflected by the 
results listed in Table \ref{limits}. About 93\% of all CP2 objects can be
detected with a limit of +10 mmag, whereas the mean value is +32.5 mmag with
an extreme value of +79 mmag. This sample includes all well-established 
CP2 stars classified
in Renson (1991) and marked with an asterisk. We have not subdivided the sample
into hotter Si and cooler CrEu(Sr) objects, because the definition is not quite
clear yet (Bychkov et al. 2003).

In Table \ref{cps} we list 296 CP2 stars which are included in Renson (1991)
but not marked as ``well established''. These objects are obviously magnetic
chemically peculiar stars because of the significant positive 
$\Delta a$ value belonging to the given subgroup. Although several
of these stars have already been assigned to the CP2 group, we see our result
as further proof of membership.

\subsection{CP3 stars}

The HgMn (CP3) stars are generally non-magnetic, slow rotating, B type stars with large
overabundances (up to five orders of magnitudes) of mercury and manganese. There are
several mechanism which play a major role in understanding these extreme peculiarities:
radiatively driven diffusion, mass loss, mixing, light induced drift, and possible 
weak magnetic fields.
However, there is no satisfactory model which explains the abundance pattern, yet 
(Adelman et al. 2003).   

\begin{figure}[t]
\begin{center}
\includegraphics[width=65mm]{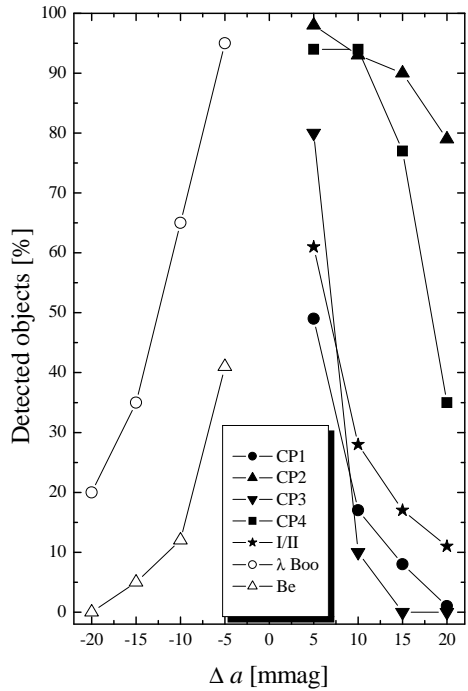}
\caption{The detection probability of the different investigated groups
as listed in Table \ref{limits}.}
\label{glimits}
\end{center}
\end{figure}

All CP3 stars listed by Adelman et al. (2003) with $\Delta a$ measurements (10 objects)
have been taken for our analysis. The mean value of all objects is +5.3 mmag but with
a very low maximum of +11 mmag (Table \ref{limits}). The detection limit drops from 80\% to
10\% for +5 and +10 mmag, respectively (Fig. \ref{glimits}). This is comparable to 
the values found for the
CP1 group. 

\subsection{CP4 stars}

As defined by Preston (1974), the CP4 stars comprise helium weak, B type
objects. They have strong magnetic fields (as the CP2 group) which produce
elemental surface inhomogeneities together with photometric variations. Several 
objects also show emission in the optical spectral range and signs of mass loss 
(Wahlgren \& Hubrig 2004).  
The percentage of detection (94\%) at +10 mmag is even higher than that of the CP2
group leading to the conclusion that almost all magnetic chemically peculiar
stars can be detected with the $\Delta a$ photometric system.

Within our sample, there are also four helium rich objects: HD 37017 
($\Delta a$\,=\,+2 mmag), HD 37479 (+10), HD 64740 ($-$1), and HD 209339 ($-$20).
Zboril et al. (1997) analyzed a sample of 17 helium rich objects and concluded that
several stars exhibit strong emission together with stellar activity. This might
be the reason for such a wide range of positive, as well as negative, 
$\Delta a$ values were observed similar to Be/shell stars.

\begin{figure*}[t]
\begin{center}
\includegraphics[width=175mm]{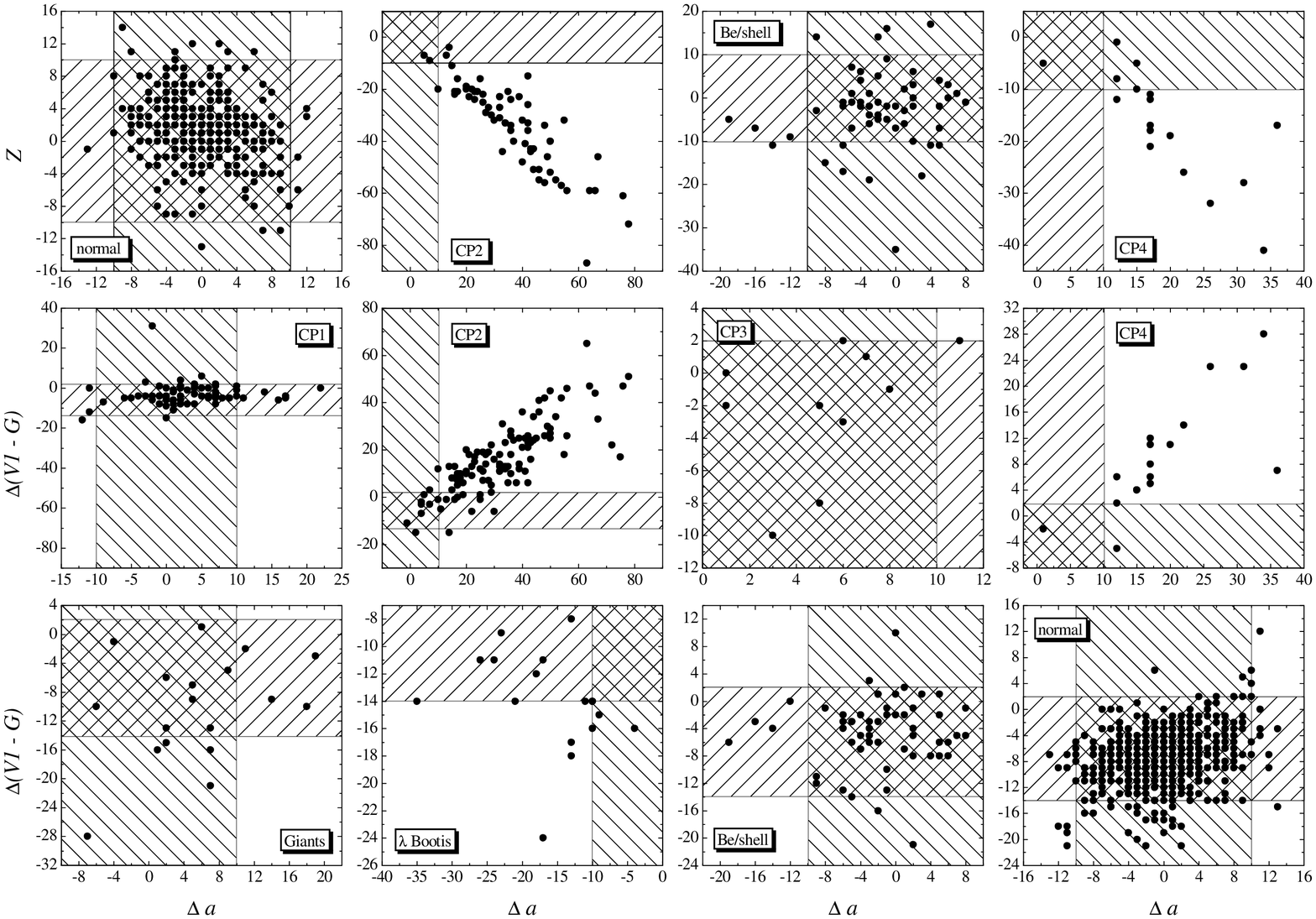}
\caption{Comparison of the detection capability of the $\Delta a$ and 
Geneva $\Delta (V1 - G)$, as well as $Z$
indices for all groups discussed in this paper. 
No objects
of the CP1 and $\lambda$ Bootis group are within 
the range of valid $Z$ values, defined for stars hotter than A0.
Because neither $Z$ nor $\Delta(V1-G)$ show significant
deviating values for CP3 objects, both diagrams are identical, so we display only
the $\Delta(V1-G)$ versus $\Delta a$ diagram. Since $Z$ is not affected by
gravity, the plot for supergiants was omitted.
Areas with the different patterns denote the regions where the respective indices
are insensitive to peculiarity. The detailed statistics of these data are listed in 
Table \ref{deltav1ga}.}
\label{deltav1g}
\end{center}
\end{figure*}

\subsection{$\lambda$ Bootis stars}

This small group comprises non-magnetic, late B- to early F-type, Population\,I,
luminosity class V
stars with apparently solar abundances of the light elements (C, N, O, and
S) and moderate to strong underabundances of Fe-peak elements (Paunzen et al. 2002a).
Only a maximum of about 2\% of all objects in the
relevant spectral domain are believed to be $\lambda$ Bootis type stars.
Two papers by Maitzen \& Pavlovski (1989a, 1989b) were dedicated to a systematical 
analysis of bona-fide group members that were selected from a catalogue by Renson et al. (1990). 
However, the group definition at this time was not clear, and
their sample included objects which definitely not belong to this group.
In our analysis, true members of this group listed by Paunzen et al. (2002a), together
with HD 84948, were included. The latter is a spectroscopic binary system which
includes two $\lambda$ Bootis type objects (Iliev et al. 2002).

The group of $\lambda$ Bootis stars is an especially excellent example of how
$\Delta a$ photometry can preselect candidates for spectroscopic observations, for
example, in young open clusters. Paunzen (2001) presents spectral classification of
708 stars selected to be good photometric candidates only on the basis of Str{\"o}mgren 
indices. From those, only 26 turned out to be new members of the $\lambda$ Bootis group.  

Within our analysis, we found twenty well-established members of the $\lambda$ Bootis group
with $\Delta a$ measurements. The group mean value is
$-$16.2 mmag, and a maximum of $-$35 mmag (Table \ref{limits})
shows the high efficiency of this photometric system. Even with a detection limit of $-$10
mmag, almost 2/3 of all bona-fide $\lambda$ Bootis stars can be detected. 

\subsection{Supergiants}

The only notice about a positive detection of supergiants was given by Vogt et al.
(1998), who investigated two cool supergiants and found a substantial positive
deviation from the normality line.

We have restricted our analysis to objects classified as luminosity class
I or II in the literature. It has to be emphasized that such objects are, in
general, easily sorted out within color-magnitude diagrams of different photometric
systems. However, Claret et al. (2003) show that isochrones with the $\Delta a$
photometric system together with the location of objects with respect to the normality
line are capable of sorting out fore- and background objects very efficiently. 

Analysis of super giants in open clusters is important in several respects. Most of these
giants are within binary systems and exhibit variability.
Furthermore, their membership is crucial for isochrone fitting because of the
sensitivity of the determined age on the existence 
of a ``giant clump'' (Eigenbrod et al. 2004).
Since all known chemically peculiar stars have luminosity classes IV or V (G\'omez et al. 1998),
super giants selected by their location in a color-magnitude diagram with a significant positive 
$\Delta a$ value can be easily tested for membership in an open cluster.
In total, eighteen supergiants from O9.5II (HD 47432) to F4Iab (HD 61715) are included in the
investigated sample with a mean value of +5.4 and a maximum value of +19 mmag 
(Table \ref{limits}).

\subsection{Be/shell stars}

These stars are defined as B type dwarfs which have shown hydrogen emission in their spectra
at least once. Due to an equatorial disk produced by stellar winds, emission arises 
quite regularly.
In addition, photometric variability on different timescales is a common phenomenon caused by
the formation of shock waves within those disks. But nonradial pulsation and variability due
to rotation are also observed (Porter \& Rivinius 2003). 

The phases of emission are replaced by shell and normal phases
of the same object. This episode was analysed for the case of Pleione using $\Delta a$ 
photometry by
Pavlovski \& Maitzen (1989). In the shell phase it reached a $\Delta a$ value of +36 
mmag, which dropped
to +4 mmag within one year. However, the behaviour of Pleione seems quite extreme and 
outstanding, because
no other similar object has been detected so far (Vogt et al. 1998). 
The contamination of classical 
chemically peculiar stars due to Be stars in a shell phase is, therefore, only marginal.

Since Pavlovski \& Maitzen (1989) already presented a paper with measurements of 
40 apparent Be/shell
type stars, our sample is rather large, 59 objects in total. The mean value is close to 
zero ($-$1.5 mmag)
with extremes of $-$19 mmag (emission phase) and +36 mmag (shell phase). 

The negative $\Delta a$ values found are probably caused by emission of iron and magnesium lines
in the spectral region from 5167 to 5197\,\AA\, (Hanuschik 1987), which fall exactly within the
$g_2$ filter and its bandwidth (Table \ref{filters}).

\subsection{Comparison with the Geneva $\Delta (V1 - G)$ and $Z$ indices}

Besides the $\Delta a$ index, the $\Delta (V1 - G)$ and $Z$ indices within the 
Geneva 7-color photometric system (Golay 1972, Cramer 1999) are the most 
suitable for detecting CP stars. The only difference between these indices is
the limitation of $Z$ to spectral types hotter than approximately A0 (Cramer 1999).
Hauck \& North (1982, 1993) investigated
the properties of $\Delta (V1 - G)$ in the context of magnetic CP stars. Here we will 
recall the definition
of $\Delta (V1 - G)$ and $Z$ as
\begin{eqnarray*}
\Delta (V1 - G) &=& (V1 - G) - 0.289\cdot(B2 - G) + 0.302 \\
Z &=& -0.4572 + 0.0255\cdot U - 0.1740\cdot B1 +  \\   
&& +0.4696\cdot B2 - 1.1205\cdot V1 + 0.7994\cdot G \\
\end{eqnarray*}
The $V1$ and $G$ filters are centered at 5408 and 5814\AA\,(bandwidths of about 
200\AA), respectively. 
The Geneva 7-color photometric system is the most homogeneous one because unique filter 
sets together with the same type of photomultipliers were used throughout its history. 
We took the sample as
described in Sect. \ref{sample} and searched for all objects with available Geneva photometry.

The zero point of the $\Delta (V1 - G)$ index represents the upper limit
of the sequence of normal type objects and not its mean value.
This was done by using the upper envelope for normal type, luminosity
class V to III objects, based
on a linear fit for the correlation of $(V1 - G)$ with $(B2 - G)$ as
given by Hauck (1974) which introduces a negative shift.
The rightmost lower panel of Fig. \ref{deltav1g} shows exactly this behavior. Only 
very few normal type stars exceed $\Delta (V1 - G)$\,$>$\,+2\,mmag, whereas
many normal type objects clearly have values lower than $-$10\,mmag with
a mean value of $-$7.6\,mmag for the complete sample. We therefore
introduced heuristic significance limits of +2 and $-$14\,mmag for $\Delta (V1 - G)$,
which brings the level of normal type objects lying outside these limits to almost
the same percentage as for the $\Delta a$ photometric system (Table \ref{deltav1ga}).
A very strict significance limit of +10\,mmag for $\Delta (V1 - G)$
was set by Hauck \& North (1982) to avoid contamination of CP objects.

The $Z$ index is virtually independent of temperature and gravity effects for
stars hotter than A0 or $(b-y)_0$\,=\,0\,mag. Cramer (1999) lists a limit of
$\pm$10\,mmag for apparent peculiarity.

Table \ref{deltav1ga} and Fig. \ref{deltav1g} show the results. The $\Delta a$ 
and Geneva 7-color 
photometric systems are able to detect magnetic CP objects (CP2 and CP4) with more
or less the same statistical significance. The slope for the CP2 stars is 0.60(6)
and a negligible zero point using only the objects, which are significant peculiar
$\Delta a$ and $\Delta (V1 - G)$ values (Fig. \ref{deltav1g}), as well as $-$0.89(7)
for $Z$, respectively. For the CP4 stars we get a slope of 1.26(14) and $-$1.31(18),
not taking the one deviating object (HD~174638) into account, respectively.

For $\lambda$ Bootis stars the detection capability is similar to that
of $\Delta a$. But for the Be/shell stars, the 
$\Delta (V1 - G)$ and $Z$ indices are even more sensitive. For
the giants, an interesting behaviour was found. While four stars exhibit significant positive
$\Delta a$ values, five stars have significant negative $\Delta (V1 - G)$ 
ones, but no positive value was found.

\begin{table}[t]
\begin{center}
\caption{A comparison of the detection capability of the $\Delta a$ 
and Geneva $\Delta (V1 - G)$, as well as $Z$
indices for the objects with available Geneva 7-color photometry. No objects
of the CP1 and $\lambda$ Bootis group are within 
the range of valid $Z$ values (defined for stars hotter than A0).
The results are shown graphically in Fig. \ref{deltav1g}.}
\label{deltav1ga}
\begin{tabular}{lrrrrr|cr}
\hline
\hline
\vspace*{-3mm}
\\
& \multicolumn{1}{c}{$N_{tot}$} & \multicolumn{2}{c}{$N_{\Delta (V1-G)}$} &
\multicolumn{2}{c|}{$N_{\Delta a}$} & \multicolumn{1}{c}{$N_{\Delta (V1-G)}$} & 
\multicolumn{1}{c}{$N_{\Delta a}$} \\
& & $-$ & + & $-$ & + & \multicolumn{1}{c}{[\%]} & \multicolumn{1}{c}{[\%]} \\
\hline
CP1 & 71 & 2 & 4 & 3 & 6 & 8 & 13 \\
CP2 & 108 & 2 & 89 & $-$ & 98 & 84 & 91 \\
CP3 & 10 & $-$ & $-$ & $-$ & 1 & $-$ & 10 \\
CP4 & 17 & $-$ & 14 & $-$ & 16 & 82 & 94 \\
I/II & 18 & 5 & $-$ & $-$ & 4 & 28 & 22 \\
$\lambda$ Boo & 17 & 11 & $-$ & 12 & $-$ & 65 & 71 \\
Be & 56 & 3 & 2 & 4 & $-$ & 9 & 7 \\
\hline
normal & 601 & 27 & 5 & 9 & 8 & 5 & 3 \\
\hline
\\
\hline
\hline
\vspace*{-3mm}
\\
& \multicolumn{1}{c}{$N_{tot}$} & \multicolumn{2}{c}{$N_{Z}$} &
\multicolumn{2}{c|}{$N_{\Delta a}$} & \multicolumn{1}{c}{$N_{Z}$} & 
\multicolumn{1}{c}{$N_{\Delta a}$} \\
& & $-$ & + & $-$ & + & \multicolumn{1}{c}{[\%]} & \multicolumn{1}{c}{[\%]} \\
\hline
CP2 & 66 & $-$ & 62 & $-$ & 63 & 94 & 95 \\
CP3 & 10 & $-$ & $-$ & $-$ & 1 & $-$ & 10 \\
CP4 & 17 & $-$ & 12 & $-$ & 16 & 71 & 94 \\
Be & 56 & 9 & 4 & 4 & $-$ & 23 & 7 \\
\hline
normal & 324 & 3 & 7 & 1 & 4 & 3 & 2 \\
\hline
\end{tabular}
\end{center}
\end{table}

\section{Conclusions}

All $\Delta a$ measurements for galactic field stars, 1474 objects in total, of the 
literature were, for the
first time, compiled and homogeneously analyzed. This intermediate band photometric system 
samples the depth of the 5200\AA\, flux depression by comparing the flux at the center
with the adjacent regions. Although it was slightly modified over the last three decades,
no systematic trend of the individual measurements was found. 
This photometric system is most suitable for detecting magnetic CP stars with high
efficiency (up to 95\% of all relevant objects). But it is also capable of detecting a 
small percentage of 
non-magnetic CP objects. Furthermore, the groups of (metal-weak) $\lambda$ Bootis, 
as well as classical Be/shell stars, can be traced with the help of this photometric system. 
In addition, we investigated the behaviour of supergiants (luminosity class I and II).
On the basis of apparent normal type objects, the correlation of the 3$\sigma$ significance 
limit and
the percentage of positive detection for all groups was derived. This is especially important
for observations in open clusters of the Milky Way and even the Magellanic Clouds.
As a next step, we compared the capability of the $\Delta a$ photometric system with the
$\Delta (V1 - G)$ and $Z$ indices of the Geneva 7-color system to detect peculiar objects. 
Both photometric systems show the same efficiency for detection of magnetic CP
and $\lambda$ Bootis stars; the indices in the Geneva system are even more efficient 
concerning the Be/shell objects.

On the basis of this statistical analysis, it is possible to derive the incidence of CP
stars in galactic open cluster and extragalactic systems, including the former unknown bias
of undetected objects.

It seems worthwhile to investigate whether
the formation of magnetic peculiar objects occurs 
in the same proportion to ``normal'' stars for all degrees of metallicity. 
Stellar models then have to explain chemically peculiar stars taking 
different metallicities, ages, and magnetic field strengths into
account as found
for different individual galactic open clusters.

\begin{acknowledgements}
We would like to thank P. North for pointing on a serious 
error in treating the Geneva measurements and for several
comments that improved this paper significant.
This work benefited from the Fonds zur F{\"o}rderung der
wissenschaftlichen Forschung, projects P17580 and P17920, as
well as the City of
Vienna (Hochschuljubil{\"a}umsstiftung project: $\Delta a$ Photometrie in der 
Milchstrasse und den Magellanschen Wolken, H-1123/2002).
Use was made of the SIMBAD database, operated at the CDS, Strasbourg, France.
This research made use of NASA's Astrophysics Data System.
\end{acknowledgements}


\begin{thebibliography}{}
\bibitem[]{} Adelman, S. J. 1980, A\&A, 89, 149
\bibitem[]{} Adelman, S. J., Adelman, A. S., Pintado, O. I. 2003, A\&A, 397, 267
\bibitem[]{} Bidelman, W. P. 1985, AJ, 90, 341
\bibitem[]{} Bidelman, W. P., \& MacConnell D. J. 1973,ApJ, 78, 687
\bibitem[]{} Bowyer, S., Sasseen, T. P., Wu, X., Lampton, M. 1995, ApJS, 96, 461
\bibitem[]{} Bychkov, V. D., Bychkova, L. V., Madej, J. 2003, A\&A, 407, 631
\bibitem[]{} Catalano, F. A., \& Leone, F. 1994, A\&AS, 108, 595
\bibitem[]{} Chen, B., Vergely, J. L., Valette, B., Carraro, G. 1998, A\&A, 336, 137
\bibitem[]{} Chu, Y.-H., Gruendl, R. A., Stockdale, C. J., Rupen, M. P., Cowan, J. J., 
Teare, S. W.
2004, AJ, 127, 2850
\bibitem[]{} Claret, A., Paunzen,E., Maitzen, H. M. 2003, A\&A, 412, 91
\bibitem[]{} Corbally, C. J. 1984, ApJS, 55, 65
\bibitem[]{} Cowley, A., Cowley, C., Jaschek, M., Jaschek C. 1969, AJ, 74, 375
\bibitem[]{} Cramer, N. 1999, New Astronomy, 43, 343
\bibitem[]{} Crawford, D. L. 1975, AJ, 80, 955
\bibitem[]{} Crawford, D. L. 1979, AJ, 84, 1858
\bibitem[]{} Crawford, D. L., \& Mander, J. 1966, AJ, 71, 114
\bibitem[]{} Deutsch, A. J. 1970, ApJ, 159, 985
\bibitem[]{} Eggen, O. J. 1980, ApJ, 238, 627 
\bibitem[]{} Eigenbrod, A., Mermilliod, J.-C., Clari{\'a}, J. J., Andersen, J., Mayor, M.
2004, A\&A, 423, 189
\bibitem[]{} Golay, M. 1972, Vistas in Astronomy, 14, 13 
\bibitem[]{} G\'omez, A. E., Luri, X, Grenier, S., Figueras, F., North, P., Royer, F., 
Torra, J., Mennessier, M. O. 1998, A\&A, 336, 953
\bibitem[]{} Gray, R. O. 1989, AJ, 98, 1049
\bibitem[]{} Hanuschik, R. W. 1987, A\&A, 173, 299
\bibitem[]{} Hauck, B. 1974, A\&A, 32, 447
\bibitem[]{} Hauck, B., \& North, P. 1982, A\&A, 114, 23
\bibitem[]{} Hauck, B., \& North, P. 1993, A\&A, 269, 403
\bibitem[]{} Hilditch, R. W., Hill, G., Barnes, J. V. 1983, MNRAS, 204, 241
\bibitem[]{} Houk, N., \& Swift, C. 1999, Michigan catalogue of two-dimensional spectral types 
for the HD Stars, Vol. 5, University of Michigan, Ann Arbor 
\bibitem[]{} Hubrig, S., Le Mignant, D., North, P., Krautter, J. 2001, A\&A, 372, 152
\bibitem[]{} Iliev, I. Kh., Paunzen, E., Barzova, I. S., Griffin, R. F., Kamp, I., 
Claret, A., Koen C. 2002, A\&A, 381, 914
\bibitem[]{} Jaschek, M., \& Jaschek, C. 1987, A\&A, 171, 380 
\bibitem[]{} Jaschek, M., Jaschek, C., Arnal, M. 1969, PASP, 81, 650  
\bibitem[]{} Kodaira, K. 1969, ApJ, 157, L59
\bibitem[]{} K{\"u}nzli, M., \& North, P. 1998, A\&A, 330, 651
\bibitem[]{} Kupka, F., Paunzen, E., Maitzen, H. M. 2003, MNRAS, 341, 849
\bibitem[]{} Maitzen, H. M. 1976, A\&A, 51, 223
\bibitem[]{} Maitzen, H. M. 1980a, A\&A, 84, L9
\bibitem[]{} Maitzen, H. M. 1980b, A\&A, 89, 230
\bibitem[]{} Maitzen, H. M. 1981, A\&A, 95, 213
\bibitem[]{} Maitzen, H. M. 1993, A\&AS, 102, 1
\bibitem[]{} Maitzen, H. M., \& Pavlovski, K. 1987, A\&A, 178, 313
\bibitem[]{} Maitzen, H. M., \& Pavlovski, K. 1989a, A\&A, 219, 253
\bibitem[]{} Maitzen, H. M., \& Pavlovski, K. 1989b, A\&AS, 81, 335
\bibitem[]{} Maitzen, H. M., \& Segewiss W. 1980, A\&A, 83, 328
\bibitem[]{} Maitzen, H. M., \& Vogt, N. 1983, A\&A, 123, 48
\bibitem[]{} Maitzen, H. M., Paunzen, E., Rode, M. 1997, A\&A, 327, 636
\bibitem[]{} Maitzen, H. M., Pressberger, R., Paunzen, E. 1998, A\&AS, 128, 537
\bibitem[]{} Maitzen, H. M., Paunzen, E., Vogt, N., Weiss, W. W. 2000, A\&A, 355, 1003
\bibitem[]{} Maitzen, H. M., Paunzen, E., Pintado, O. I. 2001, A\&A, 371, L5
\bibitem[]{} Massa, D. 1989, A\&A, 224, 131
\bibitem[]{} Maury, A. 1897, Ann. Astron. Obs. Harvard Vol. 28, Part 1
\bibitem[]{} Mermilliod, J.-C., Mermilliod, M., Hauck, B. 1997, A\&AS, 124, 349
\bibitem[]{} Michaud, G., Tarasick, D., Charland, Y., Pelletier, C. 1983, ApJ, 269, 239
\bibitem[]{} Norris, J., Bessell, M. S., Pickles, A. J. 1985, ApJS, 58, 463
\bibitem[]{} Paunzen, E. 2001, A\&A, 373, 633
\bibitem[]{} Paunzen, E., Duffee, B., Heiter, U., Kuschnig, R., Weiss, W. W. 2001, 
A\&A, 373, 625
\bibitem[]{} Paunzen, E., Iliev, I. Kh., Kamp, I., Barzova, I. 2002a, MNRAS, 336, 1030
\bibitem[]{} Paunzen, E., Pintado, O. I., Maitzen, H. M. 2002b, A\&A, 395, 823
\bibitem[]{} Pavlovski, K. \& Maitzen, H. M. 1989, A\&AS, 77, 351
\bibitem[]{} Porter, J. M., \& Rivinius, Th. 2003, PASP, 115, 1153
\bibitem[]{} Renson, P. 1991, Catalogue G\'en\'eral des Etoiles Ap et Am,
Institut d'Astrophysique Universit\'e Li\`ege, Li\`ege
\bibitem[]{} Renson, P., Faraggiana, R., B{\"o}hm, C. 1990, Bull. Inform. CDS, 38, 137
\bibitem[]{} Schnell, A., \& Maitzen, H. M. 1994, 
Proc. 25th Meeting EWG CP Szombathely, p. 87
\bibitem[]{} Schnell, A., \& Maitzen, H. M. 1995, IBVS, 4175
\bibitem[]{} Skiff, A. B. 2003, VizieR On-line Data Catalog: III/233. Originally published in: 
Lowell Observatory (2003)
\bibitem[]{} Str\"omgren, B. 1966, ARA\&A, 4, 433
\bibitem[]{} Vogt, N., Kerschbaum, F., Maitzen, H. M., Fa\'undez-Abans, M. 1998, 
A\&AS, 130, 455 
\bibitem[]{} Wahlgren, G. M., \& Hubrig, S. 2004, A\&A, 418, 1073
\bibitem[]{} Zboril, M., North, P., Glagolevskij, Yu. V., Betrix, F. 1997,
A\&A, 324, 949 
\end{thebibliography}
\end{document}